\documentclass[10pt]{article}

\usepackage{graphicx}%
\usepackage{amsmath,amssymb,amsthm,upref,bm}%

\DeclareMathAlphabet{\varmathbb}{U}{pxsyb}{m}{n}

\newcommand{\pd}[3][]{\mathchoice{\raise-0.5pt\hbox{$\partial$}%
\vphantom{\partial}_{\mkern-1.5mu#2}^{\mkern0.4mu#1}\mkern0.3mu}%
{\raise-0.5pt\hbox{$\partial$}%
\vphantom{\partial}_{\mkern-1.5mu#2}^{\mkern0.4mu#1}\mkern0.3mu}%
{\raise-0.5pt\hbox{$\scriptstyle\partial$}%
\vphantom{\partial}_{\mkern-1.7mu#2}^{\mkern0.1mu#1}\mkern0.1mu}%
{\raise-0.5pt\hbox{$\scriptscriptstyle\partial$}%
\vphantom{\partial}_{\mkern-1.7mu#2}^{\mkern0.1mu#1}\mkern0.1mu}#3}
 
\newcommand{\D}{\mathrm{d}\kern0.2pt}%
\newcommand{\ii}{\kern0.05em\mathrm{i}\kern0.05em}%
\newcommand{\E}{\textrm{e}}%
\newcommand{\RR}{\mathbb{R}}%

\begin{document}

\baselineskip=4.4mm

\makeatletter

\title{\bf A tale of two Nekrasov's integral equations}

\author{Nikolay Kuznetsov}

\date{On the occasion of centenary of Nekrasov's equation for deep water}

\maketitle

\vspace{-6mm}

\begin{center}
Laboratory for Mathematical Modelling of Wave Phenomena, \\ Institute for Problems
in Mechanical Engineering, Russian Academy of Sciences, \\ V.O., Bol'shoy pr. 61,
St. Petersburg 199178, Russian Federation \\ E-mail: nikolay.g.kuznetsov@gmail.com
\end{center}

\begin{abstract}
Just 100 years ago, Nekrasov published the widely cited paper \cite{N1}, in which he
derived the first of his two integral equations describing steady periodic waves on
the free surface of water. We examine how Nekrasov arrived at these equations and
his approach to investigating their solutions. In this connection, Nekrasov's life
after 1917 is briefly outlined, in particular, how he became a victim of Stalin's
terror. Further results concerning Nekrasov's equations and related topicz are
surveyed.
\end{abstract}

\setcounter{equation}{0}

\section{Introduction}

Theory of nonlinear water waves has its origin in the work of George Gabriel Stokes
(1819--1903) dating back to the 1847 paper \cite{S} (see also \cite{SCP},
pp.~197--219). His research in this field is well documented; see, for example, the
detailed survey \cite{Cr}, where further references are given. The next major
achievement in developing this theory was due to Aleksandr Ivanovich Nekrasov
(1883--1957), who derived two integral equations now named after him; one describes
Stokes waves on deep water, whereas the other one deals with waves on water of
finite depth. He investigated his equations by virtue of mathematical techniques
available in the 1920s, but a comprehensive theory of these and other integral
equations arising in the water-wave theory was developed much later after the
invention of abstract global bifurcation theory in the 1970s.

The pioneering paper \cite{N1}, published by Nekrasov in the Bulletin of the now
non-existent Ivanovo-Voznesensk Polytechnic Institute, is widely cited, but without
any detail about its content. The reason for this was explained by John~V. Wehausen
(1913--2005) in his re\-view of the book \cite{N3}; see MathSciNet, MR0060363. (This
memoir summarising Nekrasov's results about both integral equations was published by
the Soviet Academy of Sciences in 1951.) Wehausen writes: ``The author's work
[\dots] appeared in publications with very small distribution outside the USSR
[\dots] and consequently has not been well known.'' These publications had very
small distribution inside the USSR as well because the times when Nekrasov carried
out this research and sent it to print were extremely hectic like those described by
Charles Dickens in the opening sentence of his {\it A~Tale of Two Cities}.
\begin{quote}
It was the best of times, it was the worst of times, it was the age of wisdom, it
was the age of foolishness, it was the epoch of belief, it was the epoch of
incredulity, it was the season of Light, it was the season of Darkness, it was the
spring of hope, it was the winter of despair, we had everything before us, we had
nothing before us, we were all going direct to Heaven, we were all going direct the
other way---in short, the period was so far like the present period, that some of its
noisiest authorities insisted on its being received, for good or for evil, in the
superlative degree of comparison only.
\end{quote}

The start of 1917 was ``the season of Light'' for the privat-dozent Nekrasov who had
recently turned 33. Indeed, on 18 February 1917, just a few days before the
beginning of the February Revolution in Russia, he presented his talk ``On waves of
permanent type on the surface of a heavy fluid'' to the Moscow Physical Society. The
corresponding paper \cite{N2} was published five (!) years later; however, its value
should not be underestimated because the first step to Nekrasov's integral equation
for waves on deep water was made in it. The equation itself had appeared in
\cite{N1} in 1921, but its first widely available presentation was published by
Wehausen only in 1960 (see the extensive survey \cite{WL}, pp.~728--730).

Before that, J.\,J. Stoker (1905--1992) included references to \cite{N1} and
\cite{N3} into his classical treatise \cite{St} on water waves, just to mention that
the problem ``of two-dimensional periodic progressing waves of finite amplitude in
water of infinite depth [\dots] was first solved by Nekrassov''; see p.~522 of his
book. Shortly after that, the 4th edition of widely cited textbook \cite{MT} by
L.\,M.~Milne-Thompson (1891--1974) was published; a detailed account of Nekrasov's
equation for waves on deep water is given in it (see sect.~$14\cdot 84$), and the
papers \cite{N1} and \cite{N4} are cited, because these ``references [\dots] appear
to be useful or appropriate''. Unfortunately, some essential points that concern
Nekrasov's approach to solving his equation are omitted in \cite{MT} and an
incorrect numerical coefficient is reproduced from \cite{N3}.

The aim of the present paper is to explore how Nekrasov arrived at the equations
(named after him now) and what were his earlier results on which he built his
approach. Indeed, one has to trace Nekrasov's train of thought through his papers
\cite{N2} and \cite{N0} (sic!) that preceded \cite{N1}, because the latter provides
no clue to it containing just six references. Three of them are about exact
solutions for rotational waves (the well-known articles by F.\,J. Gerstner and
W.\,J.\,M. Rankine, and the now forgotten note \cite{G}), whereas the rest three are
as follows: the classical paper \cite{S} by Stokes, Nekrasov's own paper \cite{N2}
and Rudzki's article \cite{R}. Only the last one is of real importance for
Nekrasov's approach, but it have sunk into oblivion now. It should be also mentioned
that the final memoir \cite{N3} (its translation into English appeared in 1967 as
MRC Report no.~813 issued by the University of Wisconsin; unfortunately, it is a
rarity now) contains only one reference (see below), and so is rather useless for
our purpose.

Thus, other sources of information about Nekrasov's train of thought are needed.
Fortunately, some clues about it were provided by Leonid Nikolaevich Sretenskii
(1902--1973)---Nekrasov's colleague at Moscow University (both were professors at
the Faculty of Mechanics and Mathematics). In his book \cite{Sr} published in 1936,
Sretenskii gave many details about Nekrasov's work on water waves. Unfortunately,
these comments were omitted in the second edition published in 1977; Sretenskii did
not complete the manuscript himself and this was done by the editors after his death.

Along with a description of Nekrasov's work on water waves, his life after 1917 is
briefly outlined, in particular, to demonstrate how easy it was to become a victim
of Stalin's terror even for a prominent scientist. For further details about the
Great Terror see the book \cite{C} by R.~Conquest, originally published in 1968
under the title {\it The Great Terror: Stalin's Purge of the Thirties}.

Finally, some results from the global theory of periodic water waves are surveyed as
well as the proof of Stokes' conjectures about the wave of extreme form. (The latter
has a stagnation point at each crest, where smooth parts of the wave profile form
the angle $\pi / 3$ symmetric about the vertical line through the crest.) This
theory is based on the modern bifurcation techniques developed in the 1970s in the
framework of nonlinear functional analysis; the corresponding references can be
found, for example, in \cite{T}. In this connection, it is worth to mention the
opinion of T.~Brooke Benjamin (1929--1995) that ``[\dots] the tools available in
functional analysis can sometimes be extremely expedient in their applications to
physical problems, winning ground that is genuinely valuable by the criteria of good
science''; see \cite{B}. Indeed, in their pioneering paper \cite{KN}, Keady and
Norbury thanked Benjamin for his supposition that the formulation of Nekrasov's
equation in an operator form ``would lead to a successful existence proof'', which
was obtained by the authors with the help of the global bifurcation theory developed
for positive operators in a cone by Dancer \cite{D}.

\section{A.\,I. Nekrasov and his integral equations}

A biography of Nekrasov by J.\,J. O'Connor and E.\,F. Robertson is available online
at \newline http://mathshistory.st-andrews.ac.uk/Biographies/Nekrasov.html \newline
However, it provides rather scanty information about him and even omits the fact of
his imprisonment in 1938--1943. This may be explained by the fact that the article
is based on the note \cite{SZ} published in 1960; at that time the Soviet censorship
was active, regardless of the fact that this was the period of the so-called
Khrushchev Thaw. Furthermore, Nekrasov did not teach and undertake his ``research in
Moscow for the rest of his life'' after 1913 as is written in O'Connor and
Robertson's article. To fill in these omissions, the most essential points of his
activity after 1917 are outlined below.

\subsection{Twists and turns of Nekrasov's life after 1917}

In the book \cite{VoT} published in 2001 (it covers all aspects of Nekrasov's
biography), the authors devoted 25 pages to sketching out major points of his life;
the most interesting of them concern the Soviet period and are outlined below.

The tsarist regime was overturned in March 1917, and self-rule was granted to
universities. After that reelection of staff began because many progressive
professors had been fired by the old government. Nekrasov was reelected as a docent
by the Council of Moscow University and promoted to professorship in 1918. However,
in the fall of that year he moved to Ivanovo-Voznesensk joining the newly organized
Polytechnic Institute, and so he was on leave from Moscow University for four
years. The reason to leave was scarce food rations in overcrowded Moscow during the
Civil War, whereas the situation was much better in Ivanovo-Voznesensk due to food
supply from regions down the Volga river. The staff of the first Soviet polytechnic
(Lenin signed a decree establishing it on 6 August 1918) included many professors
evacuated from Riga in 1915 (a consequence of World War I) and several professors
from Moscow University; Luzin, Nekrasov and Khinchin were the most notable. Along
with professorship in Theoretical Mechanics, Nekrasov was Dean of the Faculty of
Civil Engineering for four years and headed the whole institute of six faculties for
13 months.

\begin{figure}[t]
\centering \includegraphics[width=0.88\linewidth]{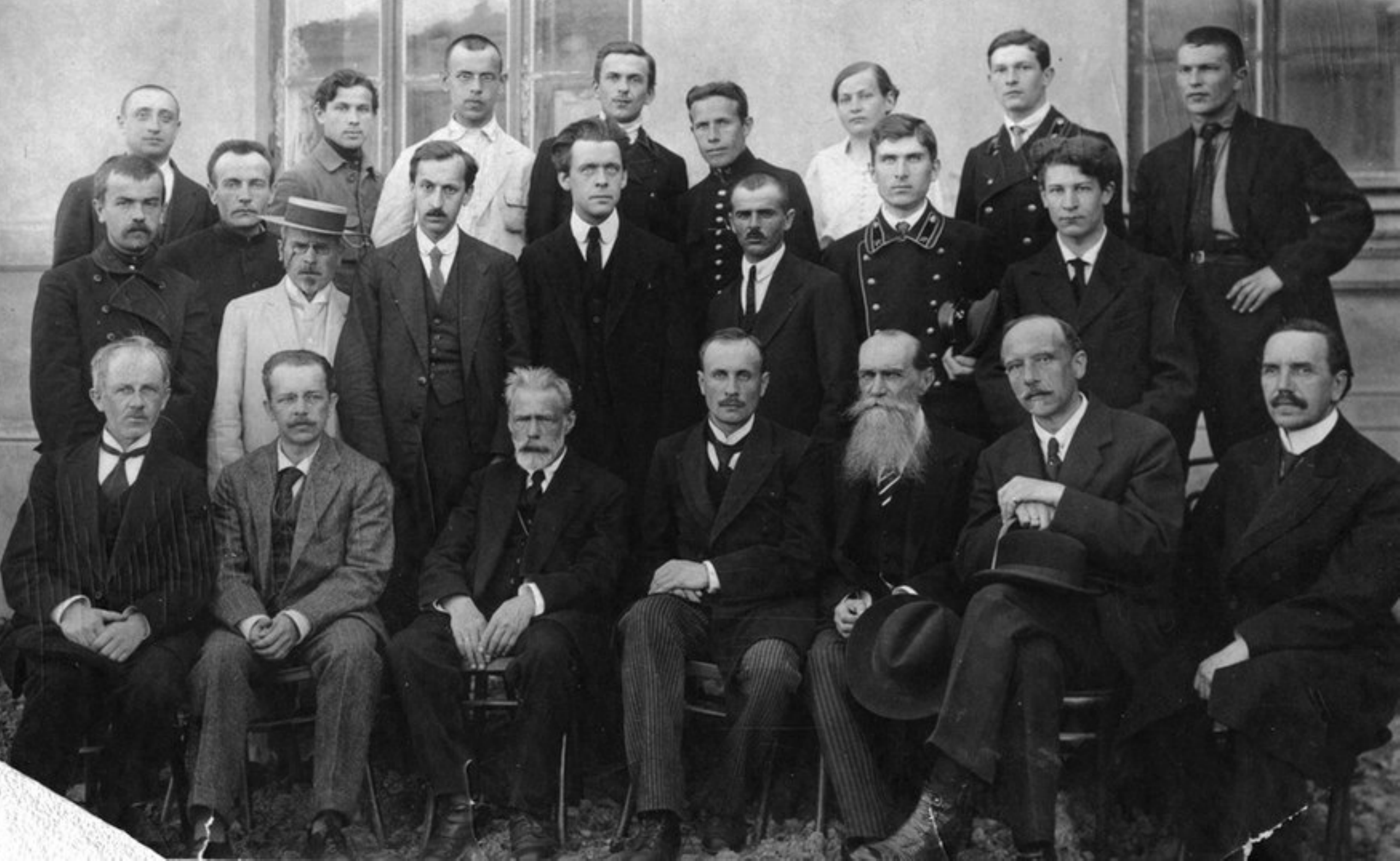} \caption{Professors
and graduate students of the Faculty of Civil Engineering, Ivanovo-Voznesensk
Polytechnic Institute. A.\,I. Nekrasov in the middle of front row.} \vspace{-4mm}
\end{figure}

He returned to Moscow University in 1922, holding simultaneously a professorship at
the Moscow Technical High School; until 1929 he combined teeching and administrative
duties at Narkompros (the Soviet ministry of education), but since that year, the
Central Aero-hydrodynamic Institute (TsAGI  founded in 1918) became his second
affiliation. There Nekrasov was drawn into mathematical problems related to aircraft
design; moreover, he was a deputy of Academician S. A. Chaplygin (1869--1942), who
headed research at the institute in the 1930s. Several times Nekrasov travelled
abroad, in particular, he was the head of the Soviet delegation at the 14th Air Show
in Paris in 1934. Next year he spent six months in the USA together with A. N.
Tupolev (1888--1972)---a prominent aircraft designer---and other researches from
TsAGI; their aim was to overview aircraft production and commercial operation of
airlines in the USA. This was a period of intensive collaboration between the
Soviets and Americans; on the one hand, the Soviet industry was interested in
American technologies, whereas selling production to the Soviet Union helped
American corporations to recover after the Great Depression.

\begin{figure}[t]
\centering \includegraphics[width=0.8\linewidth]{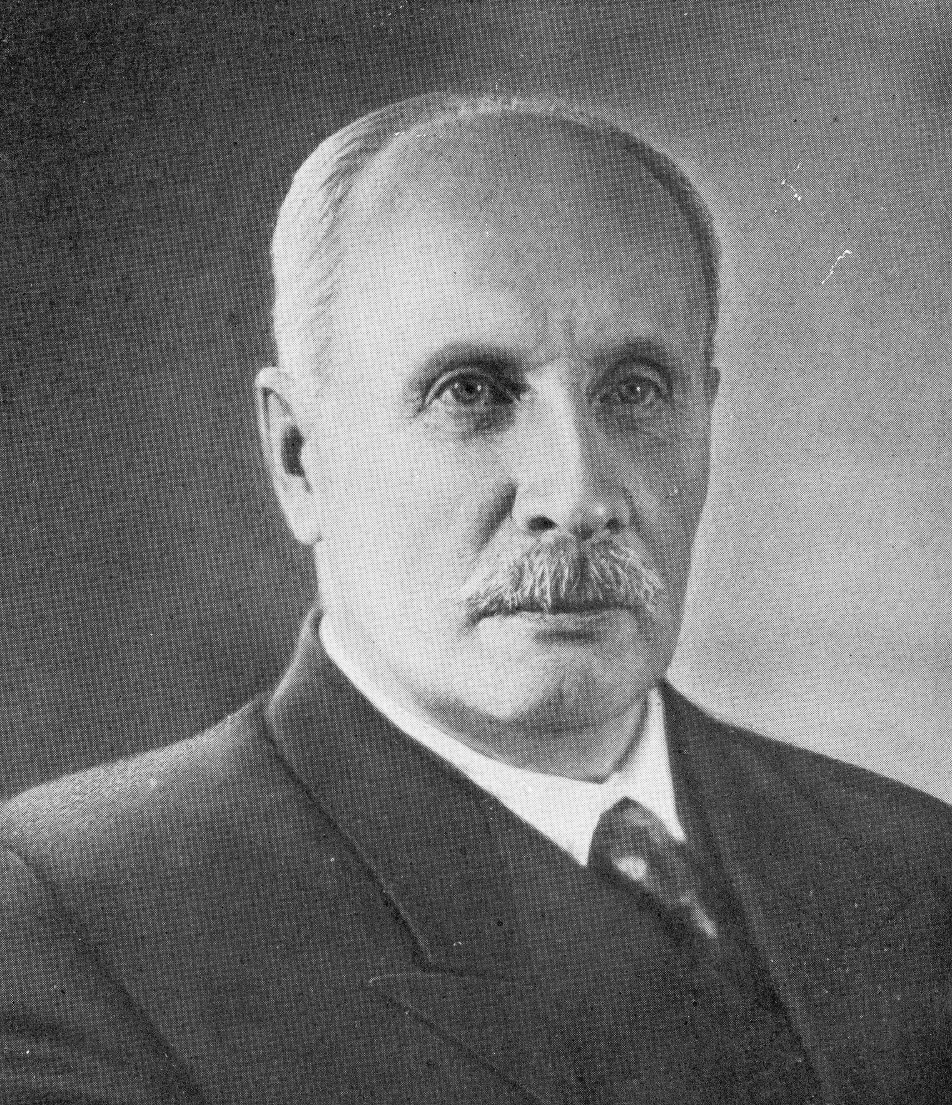} \caption{A.\,I.
Nekrasov in the 1950s.} \vspace{-4mm}
\end{figure}

Tupolev's design office separated from TsAGI in 1936, but ceased its existence next
year when Tupolev and many of his colleagues were arrested during the Great Terror
organised by the People's Commissariat for Internal Affairs (the interior ministry
of the Soviet Union) known as NKVD. Nekrasov's turn came in 1938, when he was
charged with ``participation in anti-soviet, spy organisation in TsAGI''.  He spent
next five years behind the bars in Tupolev's ``sharashka'' (this is the argot term
for ``Special Design Bureau at the NKVD''). This ``bureau'' was one of hundreds
existed within the special department created in 1938 by Beria (he headed NKVD at
that time) and disbanded in 1953, shortly after Stalin's death. (For general
information about sharashkas see, for example, the Wikipedia article available at
https://en.wikipedia.org/wiki/Sharashka. Tupolev's sharashka is described in detail
by L.~L. Kerber \cite{Ker}, who was an aviation specialist and had the long
professional and personal relationship with Tupolev; see also \cite{Jo} for a review
of \cite{Ker}, where some omissions are mentioned.)

Nekrasov was released in 1943 (exonerated only in 1955, two years before his death),
and again he returned to Moscow University, holding simultaneously a position at the
Institute for Mechanics of the Soviet Academy of Sciences (he was its Corresponding
Member since 1932 and promoted to a Full Academician in 1946). However, he continued
to head the theoretical department at Tupolev's design office until 1949. (A
characteristic feature of the time: it was located in the same sharashka's
building, but now without armed escorts.) After that, Nekrasov returned to his
studies of water waves and compiled the monograph~\cite{NCP}. His last research note
was published in 1953 and he stopped teaching the same year; the reason was asthma
contracted during imprisonment.

A few lines about Nekrasov's distinctions. In 1922 he was the first recipient of
Joukowski prize for the paper \cite{N1}, and received prize from the Narkompros for
the booklet {\it Diffusion of a vortex} ten year later. For the monograph \cite{N3}
he was awarded Stalin prize in 1952. Next year, Nekrasov was decorated with the
Order of Lenin on the occasion of his 70th birthday.

\subsection{How Nekrasov arrived at his equation for waves on deep water}

{\bf Prehistory.} In 1906, just before his graduation from Moscow University,
Nekrasov had completed a study of the motion of Jupiter's satellites for which he
was awarded a gold medal, but then he chose the water-wave theory for his further
studies. Presumably, this was due to the influence of N.\,E. Joukowski (1847--1921),
whose own research was on this theory at that time. In the paper \cite{J} of 1907,
he obtained an important result on the wave resistance of a ship; it occurred to be
similar to that published by J.\,H. Michell (1863--1940) in 1898, but Joukowski
found it independently using a different method.

In the first paper dealing with water waves (see \cite{N00}, also the first paper in
the {\it Collected Papers} \cite{NCP}), Nekrasov applied the method of power
expansions, which he knew well from his work on Jupiter's satellites and the
knowledge of which he extended while translating into Russian the 2nd volume of the
Goursat {\it Cours d'Analyse Math\'ematique} \cite{Go}. In fact, Nekrasov follows
the Stokes paper \cite{S}, where expansions in powers of the wave amplitude were
used and the latter was implicitly supposed to be a small parameter. On the
contrary, such a parameter is introduced by Nekrasov explicitly into the nonlinear
Cauchy--Poisson problem as it is referred to now; namely, the motion is generated by
an impulsive pressure applied to the surface of initially resting fluid. Assuming
the pressure impulse given by a converging series in powers of the parameter with
the zero initial coefficient, Nekrasov derives the initial-boundary value problems
for four initial harmonic functions in the expansion of the velocity potential and
finds particular solutions for two of these problems.

In his second paper on water waves (we recall that \cite{N2} was published with a
long delay), Nekrasov radically changes the topic of research turning to the
two-dimensional problem of steady gravity waves on deep water. This problem was 70
years old at that time, but he developed a new approach to it. First, he formulated
it in terms of the {\it complex velocity potential} $w (z) = \varphi + \ii \psi$
considered as a function of the complex variable $z = x + \ii y$ in the domain under
the wave profile $y = \eta (x)$. (It should be noticed that it was Michell, who
first used methods of complex analysis in the water-wave theory; see his pioneering
paper \cite{M} on the Stokes wave of the extreme form.) Second, Nekrasov proposed a
new transformation of the problem which occurred one of the key points on the way to
deriving the integral equation.

In our presentation of Nekrasov's results, his terminology and notation varying from
paper to paper is unified and updated; the velocity field is $\nabla \varphi$
($\nabla$ denotes the gradient) and so on (see, for example, \cite{T} or \cite{BT},
ch.~10, for the detailed statement of the problem).

Nekrasov assumed that the fluid is infinitely deep and
\begin{equation}
\lim_{y \to -\infty} \nabla \varphi = (c, 0) \, , \label{c}
\end{equation}
whereas waves are periodic and symmetric about their troughs and crests and their
wavelength is $\lambda$. If the origin in the $z$-plane is at a trough (this is
convenient in what follows, but Nekrasov's choice of the origin was at a crest;
however, this is unimportant), then the appropriately chosen complex potential $w
(z)$ maps conformally the vertical semi-strip
\[ D_z = \{ - \lambda /2 < x < \lambda / 2 , - \infty < y < \eta (x) \}
\] 
with unknown upper boundary (it corresponds to the water domain under a single wave)
onto the fixed semi-strip
\[ D_w = \{ - c \lambda /2 < \varphi < c \lambda / 2 , - \infty < \psi < 0 \} 
\]
on the $w$-plane. The new transformation proposed in \cite{N2}, namely,
\begin{equation}
u (w) = \exp \{ 2 \pi w / (\ii c \lambda) \} \, , \label{u}
\end{equation}
maps the latter semi-strip into an auxiliary $u$-plane and the image of $D_w$ is the
unit disc cut along the nonpositive real axis:
\begin{equation}
D_u = \{ |u| < 1 ; \ r \notin (-1 , 0] \ \mbox{when} \ \theta = \pm \pi \} \, ; \ \
\mbox{here} \ u = r \E^{\ii \theta} . \label{D_u}
\end{equation} 
Then Nekrasov noticed that the mapping $D_u \to D_z$ is defined by the relation
\begin{equation}
\frac{\D z}{\D u} = \frac{\ii \lambda}{2 \pi} \frac{f (u)}{u} \, , \ \ \mbox{where}
\ f (u) = 1 + a_1 u + a_2 u^2 + \dots \ \mbox{with real coefficients} \ a_k ;
\label{Dz/Du}
\end{equation}
moreover, $f$ is analytic in the unit disc. This allowed him to derive a formula for
the potential energy in terms of $a_k$, and the following representation of the free
surface profile parametrised by $\theta \in [- \pi, \pi]$:
\begin{eqnarray}
&& \eta (\theta) = \frac{\lambda}{2 \pi} \left( a_1 \cos \theta + \frac{a_2}{2} \cos
2 \theta + \frac{a_3}{3} \cos 3 \theta + \dots \right) , \label{4} \\ && x (\theta)
= - \frac{\lambda}{2 \pi} \left( \theta + a_1 \sin \theta + \frac{a_2}{2} \sin 2
\theta + \frac{a_3}{3} \sin 3 \theta + \dots \right) . \label{5}
\end{eqnarray}
Moreover, differentiating the Bernoulli equation with respect to $\theta$,
Nek\-rasov reduced it to
\begin{equation}
\frac{\D }{\D \theta} \big[ f (\E^{\ii \theta}) f (\E^{-\ii \theta)} \big]^{-1} =
\frac{g \lambda}{\pi c^2} \frac{\D \eta}{\D \theta} (\theta) \, , \label{B}
\end{equation}
where $g$ is the acceleration due to gravity. Finally, several examples were given,
when the initial coefficients $a_k$ can be found approximately, but they are of
little interest.

In his third paper \cite{N0} (the first one published in the {\it Bulletin} of
Ivanovo-Voznesensk Polytechnic), Nekrasov extended his analysis to what he called
``the limiting form'' of periodic waves ``the possibility of which was first
predicted by Stokes'' (no reference is given). Indeed, in the 1880 collection of his
papers \cite{SCP}, Stokes published the three short appendices and 12-page
supplement to his 1847 paper \cite{S}. In the second appendix, he coined the term
``wave of greatest height'' to characterize a certain kind of periodic wave. Since
the true height $\eta_{\rm max} - \eta_{\rm min}$ may not be maximised by it, this
wave is referred to as the ``wave of extreme form'' nowadays, because there is a
$2 \pi / 3$ angle between tangents to its wave profile at a crest.

Nekrasov begins his paper with the following remark: ``a particular form of these
waves ({\it sic}\,!) was found by Michell''. Again, no reference is given, but it is
clear that this concerns the Michell's paper \cite{M}, in which a procedure for
calculating the coefficients in a series describing this wave was developed, but no
convergence was proved. Then Nekrasov formulates the aim of his paper: ``to show how
the general mode of wave can be obtained''. For this purpose he applies a
modification of the transformation used in the previous paper for smooth waves.
Instead of \eqref{Dz/Du}, the mapping $D_u \to D_z$ is described by the relation
\begin{equation}
\frac{\D z}{\D u} = \frac{-\ii \lambda}{2 \pi} \frac{\hat f (u)}{u \sqrt[3] {1-u}}
\, , \ \ \mbox{where} \ \hat f (u) = 1 + \hat a_1 u + \hat a_2 u^2 + \dots \
\mbox{with real coefficients} \ \hat a_k , \label{Dz/Du'}
\end{equation}
in order to catch its singularity. The resulting representation of the free surface
profile is as follows, $\theta \in [- \pi, \pi]$:
\begin{eqnarray}
\frac{\D \hat \eta}{\D \theta} = \frac{\lambda}{2 \pi \sqrt[3] {2 \sin \theta / 2}}
\left( \sin \frac{\pi - \theta}{6} + \sum_{k=1}^\infty a_k \sin \frac{\pi + (6k-1)
\theta}{6} \right) , \label{7} \\ \frac{\D x}{\D \theta} = \frac{\lambda}{2 \pi
\sqrt[3] {2 \sin \theta / 2}} \left( \cos \frac{\pi - \theta}{6} + \sum_{k=1}^\infty
a_k \cos \frac{\pi + (6k-1) \theta}{6} \right) . \label{8}
\end{eqnarray}
Here integration is not as simple as in the case leading to \eqref{4} and \eqref{5}.
Similar to \eqref{B}, a complicated nonlinear equation is obtained for determining
the coefficients $\hat a_k$; it equates the right-hand side term in \eqref{7} (with
an extra constant factor) and the derivative of
\[ ( \sin \theta / 2 )^{2/3} \Big/ \big[ \hat f (\E^{\ii \theta}) \hat f (\E^{-\ii
\theta}) \big] \, ,
\]
where $\hat f$ is given in \eqref{Dz/Du'}. After tedious calculations Nekrasov
obtained only two initial coefficients, but this was insufficient to achieve
Michell's accuracy. Almost 40 years later, Nekrasov's method was rediscovered by
Yamada \cite{Y}, whose calculations provided the same accuracy as Michell's.

\vspace{1mm}

\noindent {\bf The breakthrough.} In his celebrated paper \cite{N1}, Nekrasov
combined the approach developed in \cite{N2} (see \eqref{u}, \eqref{D_u} and
\eqref{B} above) and modifications of two earlier results; one obtained by M.\,P.
Rudzki (1862--1916) (the main field of his research was geophysics; see \cite{JK})
and the other one from the dissertation of N.\,N. Luzin (1883--1950) (a Soviet
Academician noted for his numerous contributions to mathematics and his famous
doctoral students at Moscow University, known as the ``Luzitania'' group). This
allowed Nekrasov to derive his integral equation for waves on deep water described
by the unknown slope of the free surface profile.

Rudzki proposed a transformation of the Bernoulli equation in 1898 (see his paper
\cite{R} and also \cite{WL}, pp.~727--728, where the essence is explained). In the
transformed equation, the sine of the angle between the velocity vector in a
two-dimensional flow and the positive $x$-axis (the function describing this angle
is harmonic) is expressed in terms of the conjugate harmonic function. This allowed
Rudzki to apply the so-called inverse procedure (see \cite{Nem} for its description)
for obtaining an exact solution for waves over a corrugated bottom; see \cite{WL},
pp.~737--739, in particular, figure~52\,b.

Nekrasov mentioned this solution in his paper \cite{N1} with a remark that Rudzki
failed to consider the case of a horizontal bottom, whereas Nekrasov's aim was to
investigate waves on deep water. Therefore, he modified Rudzki's transformation as
follows. He introduced $R (\theta)$ and $\Phi (\theta)$ so that
\begin{equation}
- \frac{2 \pi}{\lambda} \frac{\D x}{\D \theta} (\theta) = R (\theta) \cos \Phi
(\theta) \, , \quad - \frac{2 \pi}{\lambda} \frac{\D \eta}{\D \theta} (\theta) = R
(\theta) \sin \Phi (\theta) \, , \label{1.13}
\end{equation}
where $\eta$ and $x$ are defined by \eqref{4} and \eqref{5}, respectively. Thus
$\Phi (\theta)$ is the angle between the tangent to the wave profile $\eta$ and the
positive $x$-axis, and it is parametrised by $\theta \in [-\pi, \pi]$ over a single
period between troughs. In new variables, the differentiated Bernoulli equation
\eqref{B} takes the form
\begin{equation}
\frac{\D  R^{-3}}{\D \theta} = \frac{3 \, g \lambda}{2 \pi c^2} \sin \Phi = \frac{\D
\exp \{ - 3 \log R \}}{\D \theta} \, , \label{9'} 
\end{equation}
which, as Nekrasov emphasised in his paper, was a crucial point of his
considerations. It is worth mentioning that formula (32.88) in Wehausen's
description of Rudzki's transformation (see \cite{WL}, p.~728) is similar to the
second equality \eqref{9'}.
 
The last but not least point in deriving an equation for $\Phi$ was to deduce
another relation involving $\Phi$ and $R$ simultaneously. It was Luzin---a
Nekrasov's colleague at Ivanovo-Voz\-ne\-sensk Polytechnic---who pointed the way to
obtaining such a relation (this is acknowledged in a footnote in \cite{N1}). In his
outstanding dissertation \cite{L} published in 1915, Luzin proved the following
theorem, which involves the singular integral operator $\mathcal C$ defined almost
everywhere on $[-\pi, \pi]$ by the formula
\[ \mathcal C U (\theta) = \frac{1}{2 \pi} P V \!\!\! \int_{-\pi}^\pi \! U (\tau) 
\cot \frac{\theta - \tau}{2} \, \D \tau \, , \ \ \mbox{where} \ P V \ \mbox{stands
for the Cauchy principal value} .
\]
{\it Let $U$ be a function harmonic in the open unit disc $\mathcal D$ centred at
the origin. If
\[ U (\theta) = \lim_{r \to 1} U (r \E^{\ii \theta}) \ belongs \ to \ L^2 (-\pi, \pi),
\]
then $V (\theta) = \mathcal C U (\theta)$ is also in $L^2 (-\pi, \pi)$ and provides
boundary values of $V\!$---the harmonic conjugate to $U$ in $\mathcal D$.}

This theorem provoked numerous generalisations (see, for example, \cite{Z} for a
description), the first of which was due to Privalov (a member of ``Luzitania''),
who demonstrated \cite{P} that for all $\alpha \in (0, 1)$ the operator $\mathcal C:
C^\alpha [-\pi, \pi] \to C^\alpha [-\pi, \pi]$ is bounded. Another result from the
same area of harmonic analysis as Luzin's theorem is the so-called Dini's formula
(see, for example, \cite{H}, pp.~266--267). Presumably, Luzin realised that a
consequence of this formula would provide a required relation connecting $\Phi$ and
$R$ and recommended Nekrasov to consider this option. Indeed, the following
corollary of Dini's formula was obtained in~\cite{N1}.

{\it Let $U + \ii V$ be holomorphic in $\mathcal D$, and let $V (\theta) = \lim_{r
\to 1} V (r \E^{\ii \theta})$ be absolutely continuous on $[-\pi, \pi]$ and such
that $V (2 \pi - \theta) = V (\theta)$ for all $\theta \in [-\pi, \pi]$. Then}
\begin{equation}
U (\theta) = \frac{1}{2 \pi} \int_{-\pi}^\pi \! V' (\tau) \log \left| \frac{\sin
(\theta + \tau) / 2}{\sin (\theta - \tau) / 2} \right| \D \tau + \mathrm{const} \, .
\label{10}
\end{equation}
Two formulae similar to \eqref{10}, but expressing $V (\theta)$ in terms of $U'
(\theta)$ were also given in \cite{N1} for symmetric and antisymmetric $U (\theta)$.

Then Nekrasov applied his proposition to the function $\ii \log f (u)$; it is
holomorphic in~$\mathcal D$, has $U (\theta) = - \Phi (\theta)$ and $V (\theta) =
\log R (\theta)$, and the last function satisfies the symmetry condition. It is
clear that \eqref{10} turns into
\begin{equation}
\Phi (\theta) = \frac{1}{2 \pi} \int_{-\pi}^\pi \frac{\D \log R}{\D \tau} \log
\left| \frac{\sin (\theta - \tau) / 2}{\sin (\theta + \tau) / 2} \right| \D \tau 
\label{11}
\end{equation}
in this case; here the constant vanishes because $\Phi (0) = 0$ for a symmetric
wave. Excluding $R$ from this relation by virtue of \eqref{9'}, Nekrasov arrived at
his integral equation for waves on deep water which are symmetric about the vertical
through a crest (and trough):
\begin{equation}
\Phi (\theta) = \frac{\mu}{6 \pi} \int_{-\pi}^\pi \frac{\sin \Phi (\tau)}{1 + \mu
\int_0^\tau \sin \Phi (\zeta) \, \D \zeta} \log \left| \frac{\sin (\theta + \tau) /
2}{\sin (\theta - \tau) / 2} \right| \D \tau \, , \ \ \theta \in [-\pi, \pi] \, .
\label{14}
\end{equation}
Here $\mu$ is a non-dimensional parameter which, first, arose as the constant of
integration while determining $R$ from \eqref{9'}. Subsequently, Nekrasov found an
expression for this parameter in terms of characteristics of the wave train;
namely:
\begin{equation}
\mu = \frac{3}{2 \pi} \frac{g c \lambda}{q_0^3} \, . \label{mu}
\end{equation}
Here $g$, $c$ and $\lambda$ were defined above, whereas $q_0$ is the velocity at a
crest assumed to be non-zero.

What else can be found in the paper \cite{N1}? First, a restriction on $\mu$ is
obtained. Assuming that $|\Phi (\theta)|$ of a solution to \eqref{14} is bounded by
$M$, Nekrasov demonstrated that the corresponding $\mu$ must satisfy the inequality
\begin{equation}
\mu < \left[ \pi M + \frac{\sin M}{3 M} \right]^{-1} . \label{4.5}
\end{equation}
Second, it is demonstrated that {\it the equality $\Phi (\theta) = \Phi (\pi -
\theta)$ cannot hold for all $\theta \in (0, \pi/2)$, if $\Phi$ is a nontrivial
solution of} \eqref{14}. This proves rigorously the observation made by Stokes that
his approximate wave profile has sharpened crests and flattened troughs; see the
corresponding figure in \cite{SCP}, p.~212, reproduced in \cite{Cr}, p.~31.

Finally, several pages near the end of~\cite{N1} are concerned with various
modifications and simplifications of equation \eqref{14} obtained under different
assumptions about the wave type (gentle slope etc.). In the last paragraph, Nekrasov
introduced the linearisation of \eqref{14}:
\begin{equation}
\Phi (\theta) = \frac{\mu}{6 \pi} \int_{-\pi}^\pi \Phi (\tau) \, \log \left|
\frac{\sin (\theta + \tau) / 2}{\sin (\theta - \tau) / 2} \right| \D \tau \, . \ \
\theta \in [-\pi, \pi] \, , \label{L}
\end{equation}
Moreover, he claimed that the set of its characteristic values is $\{3,6,9,\dots\}$,
whereas the corresponding eigenfunctions are $\{\sin \theta, \sin 2 \theta, \sin 3
\theta, \dots\}$.

\subsection{Nekrasov's approach to the existence of nontrivial solutions}

At the end of his paper \cite{N1}, Nekrasov announced that a method of solution of
the integral equation \eqref{14} would appear in a separate paper; its manuscript
was prepared under the title ``On steady waves, part~3'', but never published as
such. According to the concluding remark in \cite{N1}, the method is based on the
theory developed for a certain class of nonlinear integral equations; the latter was
presented in \cite{N4} (the last Nekrasov's publication in the {\it Bulletin} of
Ivanovo-Voznesensk Polytechnic, 1922). The unpublished manuscript was used while
preparing \cite{N3} for publication almost 30 years later, and in the course of that
some improvements were proposed by Y.\,I. Sekerzh-Zenkovich (1899--1985) (each
acknowledged in the text). Subsequently, he became the editor of Nekrasov's {\it
Collected Papers}, I, \cite{NCP}, and translated \cite{MT} and \cite{St} into
Russian.

\begin{figure}
\centering \includegraphics[width=0.9\linewidth]{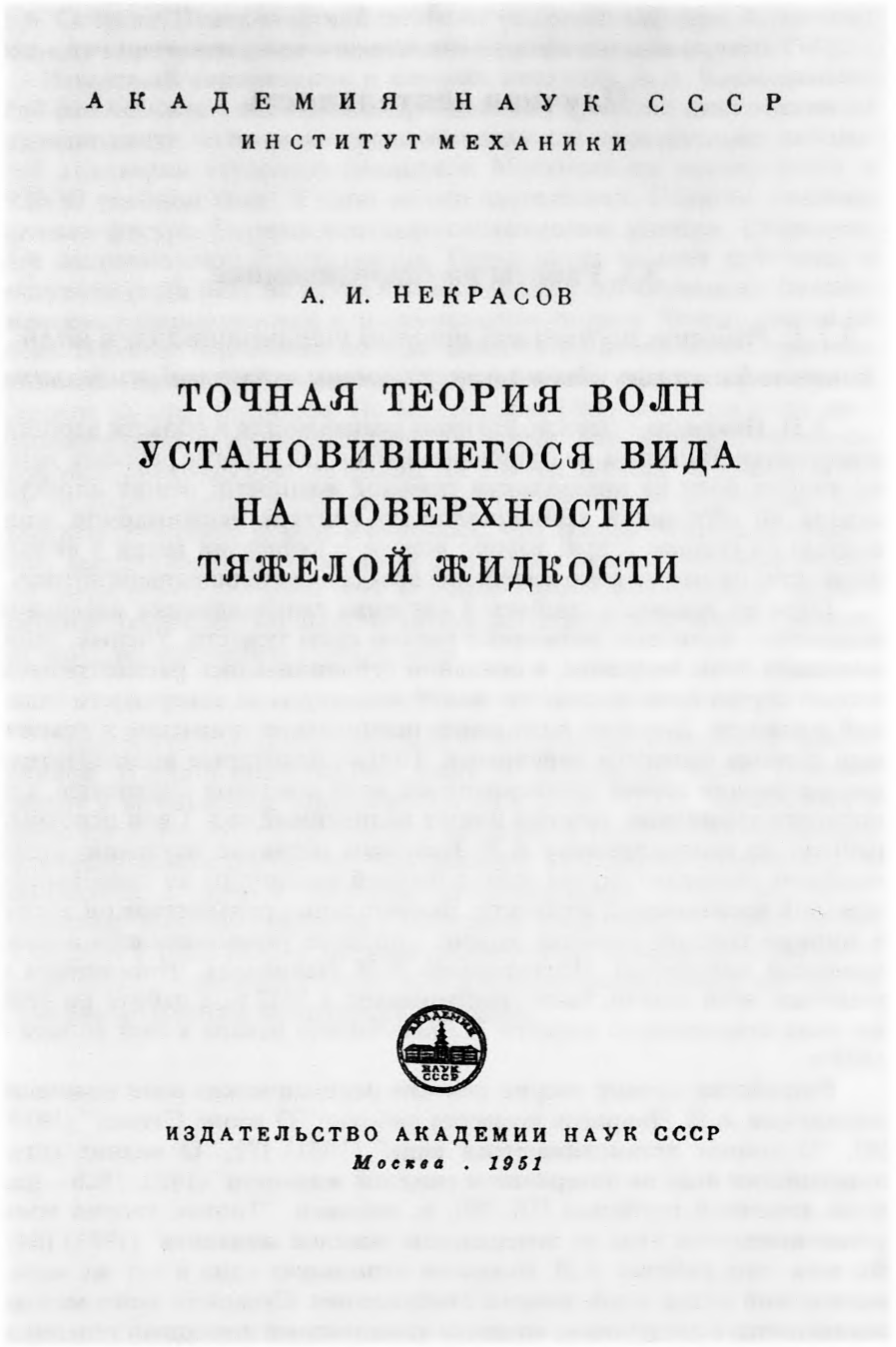} \caption{The title
page of Nekrasov's monograph \cite{N3}.} \vspace{-2mm}
\end{figure}

Let us outline key points of Nekrasov's approach to the existence of nontrivial
solutions as they are presented in \cite{N3}, mainly in Sections~5 and 6. First, he
used a straightforward calculation based on Euler's formula to transform the kernel
in equation \eqref{14}, namely:
\[ \log \left| \frac{\sin (\theta + \tau) / 2}{\sin (\theta - \tau) / 2} \right| =
2 \sum_{k=1}^\infty \frac{\sin k \theta \sin k \tau}{k} \, .
\]
This sum divided by $3 \pi$ was denoted by $K (\theta, \tau)$ and equation
\eqref{14} written in the form:
\begin{equation}
\Phi (\theta) = \mu \int_{-\pi}^\pi \frac{\sin \Phi (\tau)}{1 + \mu \int_0^\tau \sin
\Phi (\zeta) \, \D \zeta} \, K (\theta, \tau) \, \D \tau \, , \ \ \theta \in [-\pi,
\pi] \, . \label{5.3}
\end{equation}
Moreover, Nekrasov noticed that the assertion about the set of solutions of the
linearised equation \eqref{L} immediately follows from the expression for $K
(\theta, \tau)$. 

Since the right-hand side in inequality \eqref{4.5} is equal to $3$ when $M=0$, he
concluded that there are no non-trivial solutions when $\mu \leq 3$. Therefore, it
is reasonable to introduce $\mu' = 3 - \mu > 0$ and to seek a solution in the form
\[ \Phi (\theta, \mu') = \mu' \Phi_1 (\theta) + \mu'^2 \Phi_2 (\theta) + \mu'^3 \Phi_3
(\theta) + \mu'^4 \Phi_4 (\theta) + \dots \, ,
\]
thus reducing \eqref{5.3} to an infinite system of equations for $\Phi_k$. Nekrasov
also emphasised that the series would represent a solution only if its convergence
is proved. In modern terms, his aim was to construct an initial part of the
solution branch bifurcating from the first characteristic value of the linearised
operator.

Another Nekrasov's aim was to obtain $\Phi_1$, $\Phi_2$ and $\Phi_3$ explicitly, and
so he noticed that $\sin \Phi = \mu' \Phi_1 + \mu'^2 \Phi_2 + \mu'^3 (\Phi_3 -
\Phi_1^3 / 6) + \mu'^4 (\Phi_4 - \Phi_1^2 \Phi_2 / 2) + \dots$ and
\[ \frac{(3+\mu') \sin \Phi (\tau)}{1 + (3+\mu') \int_0^\tau \sin \Phi (\zeta) \, 
\D \zeta} = \mu' 3 \Phi_1 + \mu'^2 \Big[ 3 \Phi_2 + \Phi_1 - 9 \, \Phi_1 \!
\int_0^\tau \!\! \Phi (\zeta) \, \D \zeta \Big] + \dots \, .
\]
The last expansion was written down by Nekrasov with four terms as well, but this
took three extra lines plus three lines of additional notation. Substituting these
expansions into \eqref{5.3} and equating the coefficients at every $\mu'^k$
($k=1,2,\dots$), he would obtain a recurrent sequence of linear integral equations
for $\Phi_k$ with free terms depending on solutions of previous ones. The first two
equations are as follows:
\begin{eqnarray}
&& \Phi_1 (\theta) = 3 \int_{-\pi}^\pi \Phi_1 (\tau) \, K (\theta, \tau) \, \D \tau
\, , \quad \theta \in [-\pi, \pi] \, , \label{5.8} \\ && \Phi_2 (\theta) = 3
\int_{-\pi}^\pi \! \Phi_2 (\tau) \, K (\theta, \tau) \, \D \tau  \label{5.8'} \\ &&
\ \ \ \ \ \ \ \ + \int_{-\pi}^\pi \! \Phi_1 (\tau) \Big[ 1 - 9 \! \int_0^\tau \!\!
\Phi_1 (\zeta) \, \D \zeta \Big] K (\theta, \tau) \, \D \tau \, . \nonumber
\end{eqnarray}
For obtaining $\Phi_1$, $\Phi_2$ and $\Phi_3$ two more equations are needed and the
corresponding formulae are given in \cite{N3}; they look awful, but are still
tractable.

It is clear that a non-trivial solution of \eqref{5.8} is $\Phi_1 (\theta) = C_1
\sin \theta$ with $C_1 \neq 0$ to be found from the orthogonality of this function
and the free term in \eqref{5.8'}---the condition of solubility of this equation.
The corresponding value is $C_1 = 1/9$, and so, after an obvious simplification,
\eqref{5.8'} takes the form:
\begin{equation}
\Phi_2 (\theta) = 3 \int_{-\pi}^\pi \! \Phi_2 (\tau) \, K (\theta, \tau) \, \D \tau
+ 108^{-1} \sin 2 \theta \, , \quad \theta \in [-\pi, \pi] \, . \label{5.9}
\end{equation}
Hence $\Phi_2 (\theta) = C_2 \sin \theta + 54^{-1} \sin 2 \theta$, and the same
procedure as above yields $C_2 = - 8 / 243$. After determining $\Phi_3 (\theta)$,
Nekrasov concluded Section~5 with the formula
\begin{eqnarray}
&& \!\!\!\!\!\!\!\!\!\!\!\!\!\!\!\!\!\!\!\!\!\! \Phi (\theta, \mu') = \Big(
\frac{1}{9} \mu' - \frac{8}{243} \mu'^2 + \frac{115}{13122} \mu'^3 + \dots \Big)
\sin \theta \nonumber \\ && + \Big( \frac{1}{54} \mu'^2 - \frac{8}{729} \mu'^3 +
\dots \Big) \sin 2 \theta + \Big( \frac{17}{4374} \mu'^3 + \dots \Big) \sin 3 \theta
+ \dots \, , \label{5.12}
\end{eqnarray}
accompanied by the remark that it remains to snow that this series is convergent. On
the basis of this formula, Nekrasov found that the wave height is as follows:
\[ \eta_{\rm max} - \eta_{\rm min} = \eta (0) - \eta (\pi) = \frac{\lambda}{\pi}
\left[ \frac{1}{9} \mu' - \frac{8}{243} \mu'^2 + \frac{71}{6561} \mu'^3 + \dots
\right] . \]

At the beginning of the next section dealing with the question of convergence,
Nekrasov pointed out that his method of solution of nonlinear integral equations
developed in \cite{N4} (it is reproduced in \cite{N3}, Sections~13 and 14) is not
applicable directly to \eqref{5.3}. The reason is the presence of $\int_0^\tau \sin
\Phi (\zeta) \, \D \zeta$ in the integrand and to overcome this difficulty ``an
artificial trick was used in the original manuscript''. It is not clear how the
trick worked, but Nekrasov preferred to follow the method suggested by
Sekerzh-Zenkovich ``as more natural''.

Its essence is to replace \eqref{5.3} by a system of two integral equations.
Putting
\[ \Psi (\theta) = \Big[ 1 + \mu \int_0^\theta \sin \Phi (\tau) \, \D \tau \Big]^{-1}
, \ \ \mbox{one obtains} \ \ \Psi' (\theta) = - \Psi^2 (\theta) \sin \Phi (\theta)
\, ,
\]
whereas \eqref{5.3} takes the form:
\begin{equation}
\Phi (\theta) = \mu \int_{-\pi}^\pi \Psi (\tau) \sin \Phi (\tau) \, K (\theta,
\tau) \, \D \tau \, , \ \ \theta \in [-\pi, \pi] \, . \label{6.3}
\end{equation}
Since $\Psi (0) = 1$, the last differential equation yields
\begin{equation}
\Psi (\theta) = 1 - \mu \int_0^\theta \Psi^2 (\tau) \sin \Phi (\tau) \, \D \tau \, ,
\ \ \theta \in [-\pi, \pi] \, , \label{6.2}
\end{equation}
which together with \eqref{6.3} constitutes the required system for $\Phi$ and
$\Psi$. Then, as in his paper \cite{N1}, Nekrasov introduced the parameter $\mu'$,
and a solution of this system was sought in the form of the following series:
\[ \Phi (\theta, \mu') = \sum_{k=1}^\infty \mu'^k \Phi_k (\theta) \, , \quad \Psi
(\theta, \mu') = 1 + \sum_{k=1}^\infty \mu'^k \Psi_k (\theta) \, .
\]
The subsequent Nekrasov's considerations are rather vague and can hardly be
considered as a rigorous proof. This concerns equations, which form the infinite
system for $\{ \Phi_k \}$ and $\{ \Psi_k \}$, the question of solubility of this
system and, especially, Nekrasov's treatment of the method of majorising functions
that serves for proving the convergence of these series. Therefore,
Sekerzh-Zenkovich, the editor of Nekrasov's {\it Collected Papers, I}\/ (see
\cite{NCP}, where the article \cite{N3} occupies about one fifth of the content)
added five pages of comments aimed at clarifying each of these points.

Fortunately, a clear presentation of this approach is available in English; see
\cite{VT}, Section~37, where the following generalisation of equation \eqref{5.3}
\[ \Phi (\theta) = \mu \int_{-\pi}^\pi \frac{\sin \Phi (\tau) + P (\tau) \cos \Phi
(\tau)}{1 + \mu \int_0^\tau \big[ \sin \Phi (\zeta) + P (\zeta) \cos \Phi (\zeta)
\big] \, \D \zeta} \, K (\theta, \tau) \, \D \tau \, , \ \ \theta \in [-\pi, \pi]
\, ,
\]
is investigated. This equation proposed by Sekerzh-Zenkovich \cite{SZ1} describes
waves generated by a small-amplitude periodic pressure applied to the free surface
of an infinitely deep flow ($P$ is related to pressure's $x$-derivative). It is
worth noticing that $\Phi \equiv 0$ is not a solution of this equation.

Small solutions of equation \eqref{5.3} are also considered in the book \cite{VT}
(see Sections~13.4 and 13.5). The approach applied by the authors is based on the
expansion:
\[ \Big[ 1 + \mu \int_0^\theta \sin \Phi (\tau) \, \D \tau \Big]^{-1} = 
\sum_{k=0}^\infty \Big[ - \mu \int_0^\theta \sin \Phi (\tau) \, \D \tau \Big]^k .
\]
Further studies of equation \eqref{5.3} are outlined in Section 3, where methods of
nonlinear functional analysis are intensely applied; these tools were actively
developed since the 1950s.

\subsection{Nekrasov's integral equation for waves on water of finite depth}

There was a break in Nekrasov's studies of water waves owing to teaching overload,
preparing Joukowski's papers for publication after his death in 1921 and
administrative duties. This lasted until 1927, when Nekrasov presented his second
integral equation at the All-Russian Congress of Mathematicians; see the enlarged
abstract \cite{N5} reproduced (with author's permission) in the book \cite{Sr}
published in 1936.

Presentation in \cite{N5} follows that in the paper \cite{N1}, dealing with waves on
deep water, but with appropriate amendments. Namely, the stream function $\psi$ (the
imaginary part of the complex velocity potential $w$) is chosen so that it vanishes
on the free surface $y = \eta (x)$ (however, the location of the origin is not
specified), whereas the domain
\[ D_z = \{ - \lambda /2 < x < \lambda / 2 , - h < y < \eta (x) \} \, ,
\] 
corresponding to a single wave, is mapped conformally on the annular one
\begin{equation}
D_u = \{ r_0 < |u| < 1 ; \ r \notin (-r_0 , 0] \ \mbox{when} \ \theta = \pm \pi \}
\, , \ \ u = r \E^{\ii \theta} ,  \ \mbox{with} \ \log r_0 = - 2 \pi h .
\label{D_u'}
\end{equation}
Furthermore, on the basis of a formula analogous to \eqref{Dz/Du} Nekrasov concluded
that
\begin{equation}
\frac{\D w}{\D z} = \frac{- c}{\E^{\Omega (u)}} , \ \ \ \mbox{where} \ \ \Omega (u)
= a_0 + \sum_{k=1}^\infty a_k \left[ \left( \frac{u}{r_0} \right)^k + \left(
\frac{r_0}{u} \right)^k \right] \label{8.4}
\end{equation}
has real coefficients in view of the bottom boundary condition. (It should be
mentioned that the meaning of $c$, referred to as the velocity of wave at the
beginning of \cite{N5}, was cleared up only in \cite{N3}, where, at the end of
Section~8, it was demonstrated that $c$ is the mean velocity of flow at the bottom.)
For $\Phi (\theta)$ and $\Psi (\theta)$---imaginary and real part of $\Omega
(\E^{\ii \theta})$---Nekrasov obtained formulae similar to \eqref{1.13}
\[ \frac{\D x}{\D \theta} (\theta) = - \frac{\lambda}{2 \pi} \, \E^{\Psi (\theta)}
\cos \Phi (\theta) \, , \quad \frac{\D \eta}{\D \theta} (\theta) = -
\frac{\lambda}{2 \pi} \, \E^{\Psi (\theta)} \sin \Phi (\theta) \, ,
\]
which imply that $\Phi (\theta)$ is the angle between the tangent to the wave
profile $\eta$ and the positive $x$-axis. Moreover, since
\[ \Psi (\theta) = a_0 + \sum_{k=1}^\infty a_k \left( r_0^{-k}  + r_0^k \right) \cos
k \theta \, ,
\]
$a_0$ and $a_k \left( r_0^{-k}  + r_0^k \right)$ are the cosine Fourier coefficients
of $\Psi$, and so
\begin{equation}
\Omega (u) = \frac{1}{\pi} \int_{-\pi}^\pi \Psi (\theta) \left( \frac{1}{2} +
\sum_{k=1}^\infty \left[ \frac{u^k \cos k \theta}{1 + r_0^{2 k}} + \frac{r_0^{2 k}
\cos k \theta}{u^k (1 + r_0^{2 k})} \right] \right) \D \theta \, . \label{8.12}
\end{equation}
Then Nekrasov noticed that
\[ \sum_{k=1}^\infty \frac{r_0^{2 k} \cos k \theta}{u^k (1 + r_0^{2 k})} = 
\frac{u}{2} \frac{\D }{\D u} \sum_{k=1}^\infty (-1)^{k-1} \log \left[ 1 - 2
\frac{r_0^{2 k}}{u} \cos \theta + \frac{r_0^{4 k}}{u^2} \right]
\]
and a similar formula holds for the first sum in \eqref{8.12}. The sum of logarithms
is equal to the logarithm of an infinite product which is naturally representable in
terms of the Weierstrass sigma functions, whose periods $\omega$ and $\omega'$ are
such that $\omega' / \ii \omega = 4 h / \lambda$. These considerations reduced
\eqref{8.12} to the following equality
\begin{equation}
\Phi (\theta) = - \frac{1}{2 \pi} \int_{-\pi}^\pi \Psi' (\tau) \log \left|
\frac{\sigma (\omega (\theta + \tau) / \pi) \, \sigma_3 (\omega (\theta - \tau) /
\pi)}{\sigma_3 (\omega (\theta + \tau) / \pi) \, \sigma (\omega (\theta - \tau) /
\pi)} \right| \D \tau \, , \ \ \theta \in [-\pi, \pi] , \label{8.17}
\end{equation}
about which is said that it is a result of rather tedious calculations (the most
part of them is omitted in \cite{N3} as well). It should be noted that \eqref{8.17}
is analogous to \eqref{11} obtained in the case of deep water.

On the other hand, Bernoulli's equation differentiated with respect to $\theta$
yields
\begin{equation}
\frac{\D \E^{- 3 \Psi}}{\D \theta} = \frac{3 \, g \lambda}{2 \pi c^2} \sin \Phi \, ,
\ \ \mbox{and so} \ \ \Psi' (\theta) = \frac{- \mu}{3} \frac{\sin \Phi (\theta)} {1
+ \mu \int_0^\theta \sin \Phi (\tau) \, \D \tau} \, , \label{8.18}
\end{equation}
where $\theta \in [-\pi, \pi]$ and $\mu$ is the same as for deep water; see
\eqref{mu}.

Combining \eqref{8.17} and \eqref{8.18}, Nekrasov arrived at his integral equation
for waves on water of finite depth
\begin{equation}
\Phi (\theta) = \frac{\mu}{6 \pi} \int_{-\pi}^\pi \frac{\sin \Phi (\tau)}{1 + \mu
\int_0^\tau \sin \Phi (\zeta) \, \D \zeta} \log \left| \frac{\sigma (\omega (\theta
+ \tau) / \pi) \, \sigma_3 (\omega (\theta - \tau) / \pi)}{\sigma_3 (\omega (\theta
+ \tau) / \pi) \, \sigma (\omega (\theta - \tau) / \pi)} \right| \D \tau \, ,
\label{8.20}
\end{equation}
where $\theta \in [-\pi, \pi]$. This equation differs from \eqref{14} obtained for
deep water by the expression under the logarithm sign.

At the end of the note \cite{N5}, Nekrasov pointed out that
\[ \log \left| \frac{\sigma (\omega (\theta + \tau) / \pi) \, \sigma_3 (\omega 
(\theta - \tau) / \pi)}{\sigma_3 (\omega (\theta + \tau) / \pi) \, \sigma (\omega
(\theta - \tau) / \pi)} \right| = 2 \sum_{k=1}^\infty \frac{1 - r_0^{2 k}}{1 +
r_0^{2 k}} \frac{\sin k \theta \sin k \tau}{k} \, ,
\]
which was demonstrated in \cite{N3} with the help of some formulae from \cite{Ti}
(this book on elliptic functions was published in 1895, presenting the theory in an
updated for that time form). Moreover, the equality
\[ (1 - r_0^{2 k}) / (1 + r_0^{2 k}) = \tanh (2 \pi k h / \lambda)
\]
expresses the sum in terms of geometric characteristics of the wave.

Denoting by $K (\theta, \tau)$ the last sum divided by $3 \pi$, we see that equation
\eqref{8.20} takes exactly the same form as \eqref{5.3}. This similarity was
emphasised by Nekrasov and for this reason he restricted himself to calculating just
two initial terms in the expansion analogous to \eqref{5.12}, which expresses
$\Phi$---a solution of \eqref{8.20}---as a series in powers of a small parameter
$\mu'$. For $h < \infty$, he considered $\mu' = \mu_1 - \mu$, where $\mu_1$ is the
first characteristic value of $K (\theta, \tau)$; it coincides with the kernel of
the integral operator obtained by linearisation of the nonlinear one in equation
\eqref{8.20}. The set of these characteristic values, namely $\{ 3 k \coth (2 \pi k
h / \lambda) \}_{k=1}^\infty$, was announced already in \cite{N5}.

In the monograph \cite{N3}, Nekrasov concludes his considerations of equation
\eqref{8.20} with the brief Section 11, in which an important property of this
equation is formulated. Namely, since the kernel is such that $K (\theta, 2 \pi -
\tau) = - K (\theta, \tau)$, a solution satisfies the symmetry equality $\Phi (2 \pi
- \theta) = - \Phi (\theta)$, and so it is sufficient to consider an equivalent
equation on the interval $[0, \pi]$. At the beginning of Section~4, this property
was mentioned for a solution of equation \eqref{5.3} as well. However, the advantage
of reducing the equation to $[0, \pi]$ demonstrated by subsequent studies (see
below) was not used in \cite{N3}.

\vspace{1mm}

Here ends the tale about Nekrasov and his work on water waves, but this is not the
end of tale about Nekrasov's integral equations.

\section{Nekrasov's equations and functional analysis}

Along with Nekrasov's equations, which are the oldest examples of nonlinear integral
equations describing water waves, there are many others and some of them will be
mentioned below. Presumably, the first survey on this topic was published in 1964;
see \cite{Hy}.

\subsection{On local and global branches of solutions}

{\bf Existence theorems for equation (19).} It was Mark Alexandrovich Krasnosel'skii
(1920--1997) (a pioneer of nonlinear functional analysis in the Soviet Union), to
whom we owe the first consideration of equation \eqref{5.3} as an operator equation
in a Banach space. His brief note \cite{K} on this topic was published in 1956, just
five years after \cite{N3}. A general approach to bifurcation points of nonlinear
operator equations can be found in his monograph \cite{KTM}; see, in particular,
the following assertion applicable to \eqref{5.3} and \eqref{8.20}.

{\it Let $A_\mu$ $(\mu > 0)$ be a family of operators defined in a neighbourhood
of\/ $\mathbf 0$---the zero element of a Banach space $X;$ each operator is assumed
to be completely continuous and such that $A_\mu \mathbf 0 = \mathbf 0$. Let also
the Frechet derivative of $A_\mu$ be $\mu B$, where $B$ is linear (of course,
completely continuous) not depending on~$\mu$. If $\mu_0$ is an odd-multiple
characteristic value of $B$, then the equation $\Phi = A_\mu \Phi$ has a continuous
branch of nontrivial solutions $(\mu, \Phi_\mu)$ in a neighbourhood of\/ $(\mu_0,
\mathbf 0)$ and $\| \Phi_\mu \|_X \to 0$ as $\mu \to \mu_0$.}

In his note \cite{K}, Krasnosel'skii demonstrated the following about Nekrasov's
operator $A_\mu$ defined by the right-hand side in either \eqref{5.3} or
\eqref{8.20}. In a neighbourhood of $\mathbf 0 \in C [-\pi, \pi]$, it is completely
continuous and its Frechet derivative is $\mu B$, where
\begin{equation}
(B \phi) (\theta) = \int_{-\pi}^\pi \phi (\tau) \, K (\theta, \tau) \, \D \tau \, ,
\ \ \theta \in [-\pi, \pi] \, . \label{Fr}
\end{equation}
Since all characteristic values of $B$, namely, $\{3 k\}_{k=1}^\infty$ for infinite
depth ($\{ 3 k \coth (2 \pi k h / \lambda) \}_{k=1}^\infty$ for finite depth) are
odd-multiple, each of them is a bifurcation point of $A_\mu$; that is, in a
neighbourhood of this value equation \eqref{5.3} (\eqref{8.20}, respectively) has a
continuous branch of solutions $\Phi_\mu \in C [-\pi, \pi]$ satisfying the equation
and such that $\| \Phi_\mu \|_C \to 0$ as $\mu$ tends to the corresponding
characteristic value.

Thus, the question of local solution branches was established, but the existence of
global ones was proved for Nekrasov's equation more than 20 years later. However,
the first global result for periodic waves was proved by Krasovskii \cite{Kra} in
1961. He reduced the Levi-Civita formulation of the water-wave problem to a
nonlinear operator equation and considered it on the cone of nonnegative functions
in a Banach space. This allowed him to apply Krasnosel'skiı's theorem about positive
operators with monotonic minorants (see \cite{KTM}, Chapter~5, Section~2.6), thus
demonstrating the existence of a wave branch such that each value in $(0, \pi / 6)$
serves as $\max \Phi$ for some wave profile belonging to the branch.

Further story is about the existence of a global branch of solutions to Nekrasov's
equation. The abstract bifurcation theory in the form developed by Rabinowitz
\cite{Rab} and Dancer \cite{D} in the 1970s provided tools for proving this result.
A self-contained analysis by Toland (see \cite{T}, in particular Sections~8 and 9)
presents the reasoning of Keady and Norbury \cite{KN}, who obtained a {\it continuum}
of solutions, that is, a maximal closed connected set of them.

Their result is formulated in terms of a cone in a real Banach space. Since the
origin is chosen at a trough, it is reasonable to use the subspace of $C [-\pi,
\pi]$ that consists of odd functions vanishing at zero and $\pi$. The closed convex
cone, say $\mathcal{K}$, in this subspace consists of functions satisfying: (i) $f
(t) \geq 0$ for $t \in [0, \pi]$; (ii) $f (t) / \sin (t/2)$ is nonincreasing on $[0,
\pi]$; (iii) $f (t) \leq f (s)$ for all $t \in [\pi/2, \pi]$ and $s \in [\pi - t,
t]$. The crucial point is that the operator $B$ maps the cone of functions
nonnegative on $[0, \pi]$ into itself as well as $\mathcal{K}$ to $\mathcal{K}$. Now
we are in a position to formulate the following theorem.

{\it An unbounded continuum $C$ of solutions of equation \eqref{14} exists in $[0,
\infty) \times \mathcal{K}$ and $(\mu, \mathbf 0)$ belongs to $C$ if and only if
$\mu = 3$. Moreover, for $(\mu, \Phi) \in C$ the following properties hold:
\begin{itemize}
\item if $\Phi$ does not vanish identically, then $\mu > 3;$
 \item $0 < \Phi < \pi/3$ and $\Phi (\theta) / \theta$ is nonincreasing on 
$(0, \pi);$ \item $\Phi' (\theta) \leq 0$ for $\theta \in [\pi/2, \pi]$.
\end{itemize}}
\noindent An interesting open question is whether there are solutions of \eqref{14}
which do not belong to $C$, but lie in a wider cone, for example, defined by
conditions (i) and (ii). From the results of Keady and Norbury it is not clear
whether $C$ is a curve as numerical computations of Craig and Nicholls demonstrate
(see Figure~1 in their article \cite{CN}); an analytical proof is still absent.

There are infinitely many continua of solutions of equation \eqref{14}. Indeed, let
$n > 1$ be an integer, then a  change of variables shows that the set
\[ \{ (n \mu, \Phi (n \theta)) : (\mu, \Phi (\theta)) \in C \}
\]
is also a continuum of solutions; it bifurcates from $(3 n, 0)$ and each element of
this continuum yields a wave of minimal period $\lambda /n$.

To the best author's knowledge there are no analogous existence results for equation
\eqref{8.20} describing waves on water of finite depth. However, another integral
equation for this case was investigated by Norbury \cite{Nor}. Since it arises after
the hodograph transform, its unknown is the angle of the flow velocity with the
horizontal as a function of the potential along the free surface profile. The
corresponding nonlinear operator is positive and completely continuous on continuous
functions, and so the same approach as in \cite{KN} yields an unbounded continuum of
solutions; moreover, it includes those found in \cite{Kra}.

\vspace{1mm}

{\bf Properties of solutions belonging to C.} Let $\Phi_\mu \in C$ solve equation
\eqref{14} for $\mu > 3$, then this non-vanishing identically function has several
interesting properties along with those listed above. First, McLeod \cite{ML}
demonstrated that {\it for $\mu$ sufficiently large} $\| \Phi_\mu \| > \pi / 6$.
Here and below $\| \cdot \|$ stands for the standard norm in $C [0, \pi]$.
Subsequently, Amick \cite{A} obtained an upper bound of this norm; combining its
slightly improved version (see \cite{To}, p.~36) and McLeod's result, one has
\begin{equation}
\pi / 6 < \sup \{ \| \Phi \| : (\mu, \Phi) \in C \} < 0.5434 \approx (1.0378) \pi /
6 \, (\approx 31.13^\circ) . \label{bds}
\end{equation}
It should be mentioned that this upper bound is very close to the value 0.530
$(\approx 30.37^\circ)$ calculated numerically in \cite{LHF}. In the paper \cite{A},
Amick obtained also that $\| \Phi \| < 0.544$ for solutions of \eqref{8.20}, which
describe periodic waves on water of finite depth; by letting the wavelength
$\lambda$ to infinity this bound is extended to solitary waves.

It is clear that the following form of Nekrasov's equation for deep water
\begin{equation}
\Phi (\theta) = \int_{-\pi}^\pi \frac{\sin \Phi (\tau)}{\mu^{-1} + \int_0^\tau \sin
\Phi (\zeta) \, \D \zeta} \, K (\theta, \tau) \, \D \tau \, , \ \ \theta \in [-\pi,
\pi] \, , \label{1.1}
\end{equation}
is equivalent to \eqref{14} and \eqref{5.3}, but one can consider it with $\mu^{-1}
= 0$ (see below). Important results about $\Phi$ solving \eqref{1.1} with $\mu > 3$
were obtained by Amick and Toland; see Appendix in their paper \cite{AT2}. First, it
was shown that $\Phi$ is a real-analytic function which, in fact, is a corollary of
Lewy's theorem \cite{Le}. What is more important, the authors obtained the water
domain and the velocity field in it from the known $\Phi$; namely, the following
theorem is true.

{\it Let $\Phi$ be a solution of \eqref{1.1} with $\mu > 3$. If positive $\lambda$
and $c$ are such that
\begin{equation}
\left[ \frac{3 g \lambda}{2 \pi c^2} \right]^{1/3} \!\! = \frac{1}{2 \pi}
\int_{-\pi}^\pi \frac{\cos \Phi (\tau) \, \D \tau}{[ \mu^{-1} + \int_0^\tau \sin
\Phi (\zeta) \, \D \zeta ]^{1/3}} \, , \label{AT}
\end{equation}
then there exist the water domain $D_z$ and a complex velocity potential defined on
$\overline{D_z}$. The domain's width is $\lambda$ and it corresponds to a single
wave, whose profile is characterised locally by the angle $\Phi$. The potential is
periodic and satisfies the free-surface boundary conditions on the upper part of
$\partial D_z$, whereas $c$ is involved in the limit condition~\eqref{c} as $y \to
-\infty$.}

It is worth mentioning that a kind of explicit expression is obtained for $\eta (x)$
describing the upper boundary of $D_z$. It is rather complicated and involves two
integrals each having a variable limit of integration; one of these integrals has
the same integrand as that in \eqref{AT}.

We conclude this section with the result of Kobayashi \cite{Ko1}, which gives an
answer, at least partial, to the conjecture made by Toland at the end of his paper
\cite{T}; it says that $C$ is a curve. Indeed, Kobayashi established the uniqueness
theorem for solutions of equation \eqref{1.1} provided $\mu \in (3, 170]$, for which
purpose two different approaches were used on three subintervals. For $\mu \in (3,
3.009]$ the uniqueness was proved analytically; first it was demonstrated that it
takes place if the maximum of a certain rather complicated function is strictly less
than one. Then this inequality was obtained by virtue of tedious estimates. Two
other parameter segments --- $[3.009, 3.3]$ and $[3.3, 170]$ --- were treated with
the help of the so-called numerical verification method. However, either of these
segments requires a slightly different version of the method, and so we describe the
case when $\mu \in [3.009, 3.3]$ in order to be specific. The method's idea is to
construct a function, whose maximum is strictly less than one and which serves for
proving the uniqueness of solution by the contraction mapping principle. The
inequality was checked numerically with the rounding mode controlled.

\subsection{Nekrasov's equation for waves of extreme form on deep water}

Let us consider equation \eqref{1.1} with $\mu^{-1} = 0$ describing waves with
stagnation points at wave crests. Indeed, $\mu^{-1} = (2 \pi / 3) \, q_0^3 / (g c
\lambda)$ according to \eqref{mu}, where $q_0$ is the velocity at a crest. The
existence of an odd solution~of
\begin{equation}
\Phi^* (\theta) = \int_{-\pi}^\pi \frac{\sin \Phi^* (\tau)}{\int_0^\tau \sin \Phi^*
(\zeta) \, \D \zeta} \, K (\theta, \tau) \, \D \tau \, , \ \ \theta \in [-\pi, \pi]
\, , \label{1.2}
\end{equation}
was proved by Toland \cite{To}, who based his approach on Keady and Norbury's result
concerning the existence of solutions of \eqref{1.1} with finite $\mu$. Their method
cannot be applied directly because the operator in \eqref{1.2} is not completely
continuous unlike that in \eqref{1.1}. Therefore, Toland considered a sequence $\{
\mu_k \}$ tending to infinity and the corresponding solutions $\{ \Phi_k \}$ of
\eqref{1.1}. Since the absolute values of $\Phi_k$ are bounded, this sequence
converges weakly in $L^2$ to a nontrivial $\Phi^*$. Then Toland demonstrated that
$\Phi_k$ converge to $\Phi^*$ strongly in $L^2$, thus obtaining that $\Phi^*$ is a
solution of \eqref{1.2}; $\Phi^*$ was also shown to be continuous on $[-\pi, \pi]$,
except at zero which is a point of discontinuity, whose nature was not resolved in
\cite{To}.

This discontinuity was the topic of a conjecture made by Stokes in his
``Considerations relative to the greatest height of oscillatory irrotational waves
which can be propagated without change of form'' --- an appendix added to \cite{S}
in \cite{SCP}, pp.~225--228. Since the origin is chosen at a trough and the
wavelength is $2 \pi$, the conjecture takes the form
\begin{equation}
\lim_{\theta \uparrow \pi} \Phi^* (\theta) = \pi / 6 \, . \label{SC}
\end{equation}
In its support, a formal asymptotic argument near a stagnation point was provided
and in conclusion was added:
\begin{quote}
``[\dots] whether in the limiting form the inclination of the wave to the horizon
continually increases from the trough to the summit [\dots] is a question which I
cannot certainly decide, though I feel little doubt that .. [convexity].. represents
the truth.''
\end{quote}
Therefore, the assertion that {\it the free surface profile $\eta$ is convex between
its successive maxima} is nowadays referred to as the second Stokes conjecture about
waves of extreme form. So far it is known that there exists an extreme wave having
this property; see \cite{PT} for the proof. Its idea is to study a parametrized
family of integral equations each associated with a free-boundary problem. It occurs
that for the parameter equal to 1 the equation has a unique solution with associated
convex free boundary; moreover, a continuation argument shows that the existence and
convexity of solutions holds when the parameter goes from 1 to 1/3. Since the
equation corresponding to the latter value is \eqref{1.2}, its solution describes a
Stokes extreme wave for which the property takes place. However, this approach
leaves unanswered the question whether every extreme wave is convex between its
successive maxima.

Let us turn to conjecture \eqref{SC} which was proved in 1982, independently in
\cite{AFT} and \cite{Pl}. The authors of \cite{AFT} discussed the difficulties of
the proof (one finds its slightly amended version in \cite{T}) in the paper's
Appendix; in particular, it is said that the operator in \eqref{1.2} lacks some
properties needed for the application of fixed-point theorems used in the case of
$\mu^{-1} \neq 0$, that is, the singularity at $\tau = 0$ is absent.

To overcome these difficulties the following approximate integral equation
\begin{equation} 
\phi (\theta) = \frac{1}{3 \pi} \int_0^\infty \frac{\sin \phi (\tau)}{\int_0^\tau
\sin \phi (\zeta) \, \D \zeta} \, \log \frac{\theta + \tau}{|\theta - \tau|} \, \D
\tau \, , \ \ \theta \in (0, \infty) \, , \label{1.6}
\end{equation}
was used in \cite{AFT}. It is straightforward to check that $\phi \equiv \pi / 6$
solves it. Furthermore, it was proved that {\it this constant solution is the only
pointwise bounded solution of \eqref{1.6} such that}
\[ \inf_{\theta \in (0, \infty)} \phi (\theta) > 0 \quad and \quad \sup_{\theta \in 
(0, \infty)} \phi (\theta) \leq \pi / 3 \, .
\]
Finally, it was shown that this property of $\phi$ implies \eqref{SC} for any odd
solution $\Phi^*$ of \eqref{1.2} satisfying inequalities
\[ 0 \leq \Phi^* (\theta) < \pi /3 \ \ \mbox{for} \ \ \theta \in (-\pi, \pi) \quad 
\mbox{and} \quad \liminf_{\theta \uparrow \pi} \Phi^* (\theta) > 0 \, .
\]
Since these properties of $\Phi^*$ were established in \cite{AT2}, \eqref{SC} is
true.

The following asymptotic formula
\[  \Phi^* (\theta) = \frac{\pi}{6} + C_1 (\pi - \theta)^{\beta_1} + C_2 (\pi - 
\theta)^{2 \beta_1} + O \left( [\pi - \theta]^{3 \beta_1} \right) \quad \mbox{as} \
\theta \uparrow \pi \, ,
\]
refines the Stokes conjecture \eqref{SC}; here $\beta_1 \approx 0.802679$ is the
so-called Grant number defined as the smallest positive root of $\sqrt 3 \, (1 +
\beta) = \tan (\pi \beta / 2)$. The asymptotics was established rigorously in
\cite{AF} and \cite{ML1}; the inequality $C_1 < 0$ was proved in \cite{ML1},
whereas $C_2 > 0$ is a result of \cite{AF}.

It is worth to mention two related results. Fraenkel \cite{F} developed a
constructive approach to a variant of the limiting Nekrasov equation describing
extreme waves. Along with the existence proof, he provided a rather accurate
approximation of its solution. However, his ``results depend on the numerical
evaluation and numerical integration of functions defined by explicit formulae.
[\dots] Therefore, purists may believe that the theorems in the paper have not been
proved.'' Varvaruca \cite{V} considered a version of equation \eqref{1.2}
generalised by introducing a parameter into its right-hand side in the same way as
in the paper \cite{PT}. He derived an equivalent form of this equation involving the
$2 \pi$-periodic Hilbert transform and this allowed him to simplify the proof of the
first Stokes conjecture and to do this for its generalised form.

We conclude this section with the result of Kobayashi \cite{Ko2}, who presented a
computer-assisted proof that equation \eqref{1.2} has a unique odd solution.
Combining this fact and that of Plotnikov and Toland \cite{PT}, he concluded that
every extreme wave is convex between its successive maxima which settles the second
Stokes conjecture about these waves.

Kobayashi obtained his result in the same way as in \cite{Ko1}. He first proved that
the uni\-queness of a solution to \eqref{1.2} takes place if the supremum of a
certain rather complicated function is strictly less than one. Then this supremum
was computed numerically with the rounding mode controlled. The approximate result
0.99290443370699 guaranteed to be smaller than one, thus implying the uniqueness.

\section{Concluding remarks}

The tale of Nekrasov's integral equations would not be complete without mentioning
further development during the past forty years. Amick and Toland initiated this
development in their seminal papers \cite{AT1} and \cite{AT2} published in 1981.

Using a procedure resembling that applied by Nekrasov, an integral equation for
solitary waves was derived in \cite{AT1}; it is analogous to \eqref{5.3}, namely:
\begin{equation}
\widehat \Phi (\theta) = \mu \int_{-\pi/2}^{\pi/2} \frac{\sec \tau \sin \widehat
\Phi (\tau)}{1 + \mu \int_0^\tau \sec \zeta \sin \widehat \Phi (\zeta) \, \D \zeta} \,
\widehat K (\theta, \tau) \, \D \tau \, , \ \ \theta \in (-\pi/2, \pi/2) \, .
\label{1.17}
\end{equation}
Here, $\widehat \Phi$ denotes the angle between the real axis in the $z$-plane and
the negative velocity vector at the free surface parametrised in a special way
because it is infinite, whereas the kernel
\[ \widehat K (\theta, \tau) = \frac{1}{3 \pi} \sum_{k=1}^\infty \frac{\sin 2 k 
\theta \sin 2 k \tau}{k} 
\]
resembles that in \eqref{5.3}; finally, the equation's parameter expressed in terms
of characteristics of the flow is as follows:
\[ \mu = \frac{6 g h c}{\pi q_0^3} \, .
\]
This formula is analogous to \eqref{mu}, but involves $h$---the depth of flow at
infinity. 

Unfortunately, $\sec \tau$ is singular at $\tau = \pm \pi/2$, and so global
bifurcation theory is not applicable directly to equation \eqref{1.17}. To overcome
this difficulty Amick and Toland considered the following sequence of Nekrasov type
equations:
\begin{equation}
\widehat \Phi_n (\theta / 2) = 2 \mu_n \int_0^\pi \frac{f_n (\tau) \sin (J \widehat
\Phi_n (\tau / 2))}{1 + \mu_n \int_0^\tau f_n (\zeta) \sin (J \widehat \Phi_n (\zeta /
2)) \, \D \zeta} \, K (\theta, \tau) \, \D \tau \, , \ \ \theta \in (0, \pi) \, .
\label{1.26}
\end{equation}
Here $K$ is the same kernel as in \eqref{5.3}, $J : \RR \to \RR$ is the following
continuous function
\[ J a = \left\{ \begin{array}{ll} \ \ \, a \ {\rm if} \ |a| \leq \pi ,
 \\ \ \ \, \pi \ {\rm if} \ a \geq \pi , \\ -\pi \ {\rm if} \ a \leq -\pi ,
\end{array} \right.
\]
and the sequence of functions $\{ f_n (\tau) \}$ is defined as follows:
\[ f_n (\tau) = \left\{ \begin{array}{ll} 2^{-1} \sec (\tau / 2) \! \quad \quad \quad
\quad {\rm if} \ |\tau| \leq \pi - n^{-1} , \\ 2^{-1} \sec ([\pi - n^{-1}] / 2) \
{\rm if} \ \pi - n^{-1} \leq |\tau| \leq \pi .
\end{array} \right.
\]

It occurred that equation \eqref{1.26} is tractable in the same way as \eqref{5.3}
(see Section~3.1); that is, for every integer $n \geq 1$ there exists an unbounded,
closed and connected set $C_n$, whose elements $(\mu_n, \widehat \Phi_n)$ are
solutions of \eqref{1.26}. Furthermore, Amick and Toland demonstrated that the sets
$C_n$ converge in a certain sense to an unbounded, closed and connected set, whose
elements $(\mu, \widehat \Phi)$ are solutions of \eqref{1.17}; moreover, to any $\mu
\in [6/\pi, \infty)$ there corresponds a solution. Properties of these solutions are
investigated in detail and the existence of a solitary wave of extreme form is also
proved in \cite{AT1}.

Amick and Toland \cite{AT2} developed further the technique of Nekrasov type
integral equations proposed in \cite{AT2}; their new results were summarized as
follows:
\begin{quote}
A detailed discussion of Nekrasov's approach to the steady water-wave problems leads
to a new integral equation formulation of the periodic problem. This deve\-lopment
allows the adaptation of the methods of Amick and Toland (1981) to show the
convergence of periodic waves to solitary waves in the long-wave limit.

In addition, it is how the classical integral equation formulation due to Nekrasov
leads, via the Maximum Principle, to new results about qualitative features of
periodic waves for which there has long been a global existence theory (Krasovskii
1961, Keady \& Norbury 1978).
\end{quote}
The references mentioned here are \cite{AT1}, \cite{Kra} and \cite{KN},
respectively.

It was Konstantin Ivanovich Babenko (1919--1987), who invented an equation
alternative to Nekrasov's fpr describing steady periodic waves on deep water and
proved a local existence theorem for his equation; see the brief notes \cite{B1}
and \cite{B2} published in 1987. A detailed analysis of this quasi-linear equation
given in \cite{BDT1} and \cite{BDT2} is based on properties of the $2 \pi$-periodic
Hilbert transform involved in the equation and its variational structure; indeed, it
is the Euler--Lagrange equation of a simple functional introduced in \cite{BDT1}. A
relation between solutions of Nekrasov's and Babenko's equations established in
\cite{BT}, Section~10.4, allowed the authors to obtain some properties of the latter
equation.

Recently, it was demonstrated that the approach used in \cite{B1} yields an equation
of the same form as Babenko's in the case when water has a finite depth; see
\cite{KD}. The only difference is that the $2 \pi$-periodic Hilbert transform is
perturbed by a compact operator in the latter case, and so the existence of local
solution branches follows analogously to the case of deep water. Moreover, this
equation and its modification proposed in \cite{DK} are convenient for numerical
computation of global bifurcation diagrams, in particular, of secondary bifurcations
which may be multiple as is shown in \cite{DK}, Figure~4.

\end{document}